\documentclass[sigconf]{acmart}

\AtBeginDocument{%
  }

\setcopyright{acmlicensed}
\copyrightyear{2026}
\acmYear{2026}
\acmDOI{10.1145/XXXXXXX.XXXXXXX}
\acmConference[CHI '26]{CHI Conference on Human Factors in Computing Systems}{April 25--30, 2026}{Honolulu, HI}
\acmISBN{978-1-4503-XXXX-X/26/04}

\usepackage{longtable}
\usepackage{array}
\usepackage{ragged2e}
\usepackage{tabularx}
\usepackage{float}
\usepackage{subcaption}

\begin{document}

\title{From Answer Generators to Reasoning Facilitators: Designing AI Tutors for Mathematical Reasoning in High-Stakes Environments}

\author{Harry Feng}
\authornote{These authors contributed equally to this research.}
\email{yumingf@stanford.edu}
\affiliation{%
  \institution{Stanford University, EE}
  \city{Stanford}
  \state{CA}
  \country{USA}
}

\author{Yuan Tian}
\authornotemark[1]
\email{ytian24@stanford.edu}
\affiliation{%
  \institution{Stanford University, EALC}
  \city{Stanford}
  \state{CA}
  \country{USA}
}

\author{Erica Zhao}
\authornotemark[1]
\email{erica117@stanford.edu}
\affiliation{%
  \institution{Stanford University, MS\&E}
  \city{Stanford}
  \state{CA}
  \country{USA}
}

\renewcommand{\shortauthors}{Feng, Tian, and Zhao}

\begin{abstract}
The rapid integration of Large Language Models (LLMs) into educational technology threatens to reduce mathematical learning to mere answer generation. This paper presents a generative study, usability study, and 12-participant field deployment of AITutor, an interactive system that translates theoretical pedagogical mechanisms into concrete user interface features. We explore how junior-high students preparing for high-stakes exams (\textit{Zhongkao}) interact with AI tutoring. Through mixed-methods triangulation (7,379 telemetry events, 8 contextual observations, 10 interviews), we reveal that students actively resist traditional Socratic dialogue under time pressure, repurposing "answer-first" shortcuts as vital diagnostic checkpoints. We demonstrate how features like layered worked examples, step-linked visual grounding, and metacognitive scaffolding lower the interaction cost of reasoning repair. We contribute a "Reasoning-Centered Product Loop," offering actionable implications for designing AI that structurally supports the inspection, local repair, curriculum verification, and delayed retrieval of mathematical reasoning in the wild.
\end{abstract}



\maketitle

\section{Introduction}
The fundamental goal of mathematics education is not the generation of correct answers, but the construction and refinement of robust reasoning processes. Decades of learning-science research demonstrate that durable mathematical learning requires specific pedagogical mechanisms---such as self-explanation~\cite{Chi1989}, cognitive struggle~\cite{Kapur2008}, and metacognitive repair~\cite{Roll2011}. However, the advent of Large Language Models (LLMs) in educational technology threatens to bypass these mechanisms. By instantly delivering monolithic, highly polished solutions, modern AI tutors resolve the problem but inadvertently strip away the pedagogical scaffolding required for students to actively externalize and construct their own reasoning chains.

While classic intelligent tutoring systems (ITS) successfully supported reasoning through granular, step-based feedback~\cite{Anderson1995, VanLehn2011}, they often assumed an ideal, highly compliant learner. Our generative research reveals a sharper sociotechnical tension: the pedagogical ideal of Socratic elicitation asks students to slow down, articulate intermediate reasoning, and productively struggle, while the lived reality of China's \textit{Zhongkao} preparation pushes students toward speed, answer verification, and homework completion. Students did not reject reasoning; they rejected interfaces that made reasoning feel like a time tax before they could determine whether the answer was even relevant. Furthermore, mathematical reasoning is highly multimodal. Existing AI systems consistently fail to coordinate text with step-linked visual representations, creating extraneous cognitive load during spatial and geometric reasoning~\cite{Mayer2003, Ainsworth2006}. There remains a critical gap: How can we design AI interfaces that balance the extreme utilitarian demand for speed and answers with the slower cognitive requirements of inspection, explanation, and transfer?

To investigate this, we designed and deployed AITutor, an interactive system that translates theoretical methods for mathematical reasoning into concrete UI mechanisms. Instead of choosing between answer delivery and Socratic dialogue, AITutor treats the interface as the balancing mechanism between fast orientation and deeper reasoning. It supports reasoning through three pedagogical interventions:
\begin{itemize}
    \item \textbf{Layered Worked Examples:} Structuring solutions with an overview (knowledge tags and core analysis), sub-question orientation (labeled, expandable parts within a compound problem), expandable step-by-step explanation, a quickly accessible final-answer checkpoint tied to the explanation layers, and exam-format constraints that keep knowledge within the junior-high syllabus and methods within exam-conventional forms.
    \item \textbf{Step-Linked Visual Grounding:} Synchronizing text with dynamic diagrams to explicitly support spatial reasoning and reduce cognitive split-attention.
    \item \textbf{Metacognitive Scaffolding:} Providing contextual follow-up prompts (e.g., ``Key points,'' ``Another way'') to lower the interaction cost of repairing broken reasoning paths.
\end{itemize}

We address three research questions:
\begin{itemize}
    \item \textbf{RQ1:} How do \textit{Zhongkao} students decide when to use AI help under homework pressure?
    \item \textbf{RQ2:} What interface mechanisms make LLM-generated math solutions inspectable and repairable?
    \item \textbf{RQ3:} How does AITutor fit into authentic home-study routines over a field deployment?
\end{itemize}

Through a 12-day field study with 12 instrumented students, 10 interviews, and 8 observations, we uncover how learners repurpose AI interfaces to manage cognitive load. Our findings offer a counter-intuitive perspective on mathematical help-seeking: behaviors often viewed as shortcutting, such as checking the final answer first, frequently serve as diagnostic checkpoints that help students orient subsequent reasoning and locate specific errors. We contribute a "Reasoning-Centered Product Loop" for designing AI that moves beyond answer delivery toward reasoning inspection, local repair, curriculum verification, and delayed retrieval.

\section{Background and Related Work}

Our research sits at the intersection of intelligent tutoring systems, cognitive psychology, human-AI interaction, and existing homework-help products. To design AI interfaces that facilitate deep mathematical reasoning rather than superficial answer retrieval, we draw on five key areas of the learning sciences and a competitive review of current Chinese AI tutoring tools.

\subsection{Intelligent Tutoring Systems and Interaction Granularity}
Classic work on intelligent tutoring systems (ITS) establishes that effective tutoring requires modeling a learner's problem-solving state. Anderson et al.~\cite{Anderson1995} demonstrated that cognitive tutors succeed by exposing the goal-subgoal structure of a problem and providing immediate, context-specific feedback. While conversational approaches like AutoTutor~\cite{Graesser2005} successfully use mixed-initiative dialogue to elicit student reasoning, the effectiveness of any tutor is tightly coupled to its interaction granularity. VanLehn~\cite{VanLehn2011} argues that systems operating at the step or substep level are significantly more effective than those merely evaluating final answers. This distinction is central to LLM-based tutoring. Most current generative tools behave as \textit{answer-based} or \textit{whole-solution-based} systems: they produce a final answer or a large block of fluent text, but they do not make each reasoning move inspectable. AITutor's design goal is to use the UI to transform a powerful answer generator into a step-based reasoning environment: final answers orient the learner, but steps, substeps, diagrams, and local repair actions become the actual unit of interaction.

\subsection{Worked Examples, Feedback, and the Assistance Dilemma}
When students rely on full solutions, learning only occurs if they actively self-explain the rationale behind each step~\cite{Chi1989}. This is supported by broader work on worked examples: example-based instruction is most effective when it helps learners identify deep structure rather than copy surface procedures~\cite{Atkinson2000}, and students benefit when they explain steps in relation to domain principles~\cite{AlevenKoedinger2002}. Renkl~\cite{Renkl2002} notes that instructional explanations must be carefully structured to induce self-explanation rather than replace it; a later meta-analysis similarly finds that induced self-explanation has positive effects across learning settings~\cite{Bisra2018}. Feedback research adds a design constraint: effective feedback should be specific, manageable, and tied to the learner's current task and next move~\cite{Hattie2007,Shute2008}. Providing too much help can erode productive struggle. The "assistance dilemma"~\cite{Koedinger2007} warns that excessive scaffolding can optimize short-term performance at the expense of long-term retention. Help-seeking itself is a fragile metacognitive skill; students often avoid help or misuse it as an effort-saving shortcut~\cite{Aleven2003}. Aleven et al.'s account of help-seeking is especially useful for interface design because it treats help as having a \textit{cost}: if help is too cheap, students may overuse answers; if help is too effortful, they abandon the repair opportunity. For LLM-based tutors, the question is therefore not whether to add friction or remove friction in general. The design challenge is to make answer access fast enough for real homework pressure while making reasoning repair low-cost at the exact step where confusion occurs.

\subsection{Multimedia Learning and Multiple Representations}
In domains such as geometry, mathematical reasoning is heavily multimodal. Larkin and Simon~\cite{Larkin1987} classically argued that diagrams reduce cognitive search and computation by making spatial relations perceptually obvious. However, the benefit of diagrams is not automatic. Ainsworth's work on multiple external representations shows that representations can complement, constrain, and deepen interpretation only when learners can translate between them~\cite{Ainsworth1999,Ainsworth2006}. Mayer and Moreno~\cite{Mayer2003} highlight that text and visuals must be coordinated to reduce extraneous load; Schnotz and Bannert~\cite{Schnotz2003} further warn that poorly matched graphics can interfere with mental-model construction. For dynamic or stepwise problems, linked representations are especially important: van der Meij and de Jong~\cite{VanDerMeij2006} show that coordinated representations can support learning in simulation-based environments. If the text references an auxiliary line before the diagram updates, the resulting split-attention effect creates extraneous cognitive load that hinders spatial reasoning.

\subsection{Retrieval Practice and Delayed Learning}
While immediate help resolves current roadblocks, long-term mastery relies on retrieval practice. Studies by Roediger and Karpicke~\cite{Roediger2006, Karpicke2008} demonstrate that active retrieval significantly improves retention compared to passive rereading. Spacing matters as well: Cepeda et al.'s meta-analysis shows that distributed practice improves retention over massed practice~\cite{Cepeda2006}, and Dunlosky et al.~\cite{Dunlosky2013} identify practice testing and distributed practice as high-utility learning strategies. Transfer requires additional care. Pan and Rickard~\cite{PanRickard2018} show that test-enhanced learning can transfer, but only when later tasks require learners to reuse the relevant relation or principle. Consequently, reviewing an AI solution is insufficient; systems must translate generated solutions into delayed retrieval and structurally meaningful transfer practice.

\subsection{LLMs in Education and Calibrated Trust}
The integration of LLMs introduces a novel challenge: generative systems can appear highly fluent even when producing hallucinations or pedagogically inappropriate methods~\cite{Kasneci2023}. In educational contexts, Lee and See~\cite{Lee2004} emphasize that the goal is not maximal trust, but \textit{calibrated} trust---designing interfaces for appropriate reliance. Kizilcec~\cite{Kizilcec2016} and Bu\c{c}inca et al.~\cite{Bucinca2021} demonstrate that simply providing transparent explanations does not prevent overreliance; users must encounter cognitive forcing functions or explicit curriculum-fit evidence before accepting algorithmic output.

\subsection{Existing Products and Competitive Gap}
\begin{figure*}[t]
  \centering
  \begin{minipage}[t]{0.32\textwidth}
    \centering
    \includegraphics[width=\linewidth,height=0.18\textheight,keepaspectratio]{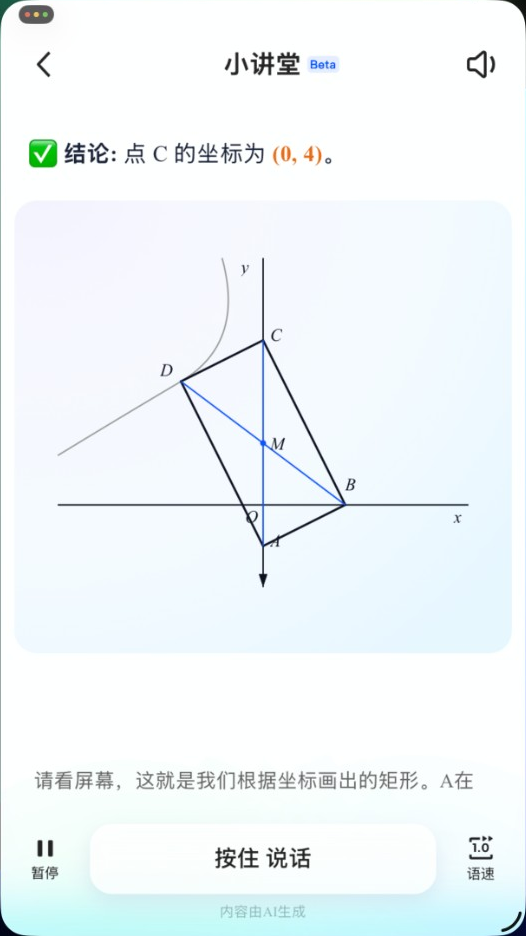}
    \caption*{\small Qianwen: synchronized lecture}
  \end{minipage}\hfill
  \begin{minipage}[t]{0.32\textwidth}
    \centering
    \includegraphics[width=\linewidth,height=0.18\textheight,keepaspectratio]{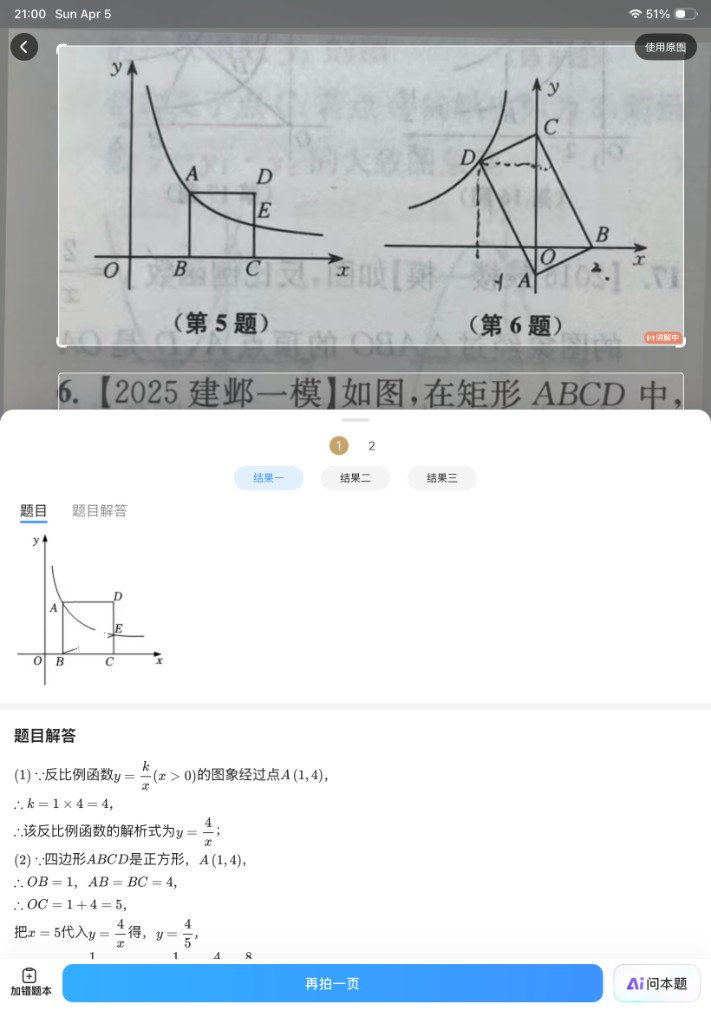}
    \caption*{\small Xiaoyuan: search mismatch}
  \end{minipage}\hfill
  \begin{minipage}[t]{0.32\textwidth}
    \centering
    \includegraphics[width=\linewidth,height=0.18\textheight,keepaspectratio]{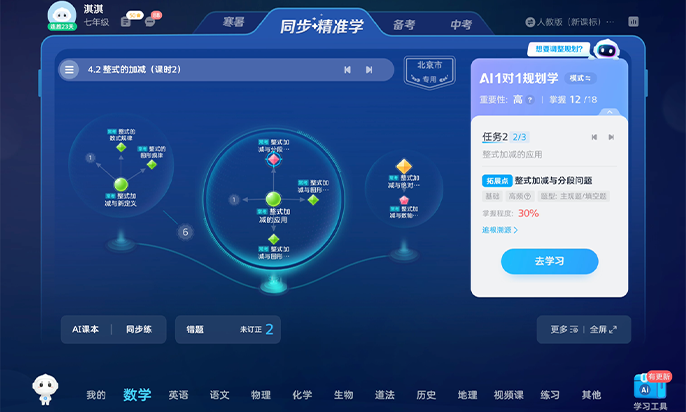}
    \caption*{\small iFlytek: diagnostic graph}
  \end{minipage}
  \caption{Competitive landscape of Chinese AI math-help products.}
  \Description{Three screenshots from existing Chinese AI tutoring tools: a Qianwen synchronized lecture screen, a Xiaoyuan mismatched search result, and an iFlytek knowledge graph dashboard.}
  \label{fig:competitive_landscape}
\end{figure*}

We also evaluated three product archetypes already available to Chinese students: Qianwen AI Tutor, Xiaoyuan AI, and the iFlytek AI Learning Tablet (Figure~\ref{fig:competitive_landscape}). This competitive review sharpened the same tension identified in the literature. Qianwen showed the promise of AI-native interaction, especially through synchronized whiteboard and voice explanation, but its general-purpose model sometimes crossed curriculum boundaries or produced exam-unsafe formats. Xiaoyuan provided a familiar zero-friction photo-search workflow, yet mismatched search results and fluent wrong explanations were displayed with the same visual authority as correct ones. iFlytek offered the strongest long-term diagnostic loop through knowledge graphs and targeted practice, but its dedicated-hardware workflow felt heavy for short homework sessions and created a cold-start burden before personalization became useful. In other words, existing products each optimize one side of the speed-reasoning tradeoff: search tools are fast but shallow, chatbots are flexible but costly to steer, and diagnostic platforms support review but feel too heavy for a late-night homework moment. Across these products, the missing pattern was consistent: students lacked a lightweight interface that could orient them quickly, ground multimodal reasoning visually, let them repair one confusing step, and verify that the method was usable in a junior-high exam.

\section{Generative Study}

To understand how junior-high students seek help in high-stakes environments, we conducted a generative study with five participants: four junior-high students (Grades 8--9) and one experienced science educator. The study combined semi-structured interviews with live walkthroughs of current homework-help tools. Our target population faces intensive math homework pressure, wrong-problem correction routines, and strict exam-oriented method constraints.

\subsection{Methods and Participants}
Participants were selected to vary by frequency of AI tool use, subject area struggles, and access to human help. Student sessions reconstructed recent stuck-problem episodes and observed how participants used existing AI or photo-search tools on real or representative problems. The educator session probed what makes an AI explanation pedagogically acceptable in an exam-oriented classroom. Together, the sessions mapped when students asked teachers versus AI, how they judged explanation quality, and what made them abandon a learning tool.

After each session, we wrote analytic memos and open-coded moments of help-seeking, answer checking, diagram use, trust breakdown, and review behavior. We then clustered repeated incidents into opportunity areas by asking what kind of interface change would reduce the observed breakdown. These opportunity areas became the five formative findings below and directly informed the first prototype.

\subsection{Participant Profiles and Contextual Workflows}
To deeply understand users' help-seeking models, we developed specific user profiles derived from our generative interviews:

\subsubsection{Amy: The Missing Step Hunter}
Amy is a Grade 9 student preparing for the Zhongkao. She generally understands class material but often gets stuck on the \textit{one} missing intermediate step in complex auxiliary-line geometry problems. She avoids asking teachers due to social friction ("I'm scared the teacher will mock me for asking a dumb question"). Amy uses AI to hunt for the specific logical bridge she missed. For her, monolithic text blocks are frustrating because they obscure the localized reasoning repair she needs.

\subsubsection{Mia: The Cautious Curriculum Validator}
Mia is a Grade 8 student with stronger academic standing. She is highly skeptical of AI in science subjects because AI tools frequently hallucinate or misinterpret diagrams. She will only invest cognitive effort into reading an explanation if she can first verify that the AI correctly extracted the geometric relationships and used methods strictly aligned with her junior-high syllabus. If the AI introduces out-of-scope concepts, she immediately abandons the session.

\subsubsection{Chloe: The Efficiency-Driven Answer Checker}
Chloe is a Grade 8 student burdened by heavy evening homework. Her primary goal is to minimize time spent on problems she already knows how to do. She approaches AI to verify her setups or diagnose simple arithmetic errors ("If I already know how to do a problem but homework isn't done yet, I want to get the answer down fast and move on"). Long, unskippable Socratic dialogue frustrates her; she demands immediate access to the final answer to decide if further reading is even necessary.

\subsubsection{Leo: The Transfer-Oriented Learner}
Leo is an ambitious Grade 8 student who wants to make sure he can solve similar problems on exams. He is not satisfied with just getting the current answer right; he explicitly checks whether he can apply the core reasoning to new contexts. However, like other students, he lacks the time to do immediate transfer practice during late-night study sessions, highlighting a need for delayed, organized review workflows.

\subsection{Key Formative Findings}
These profiles highlighted profound sociotechnical tensions between idealized learning theories and ground-level homework survival strategies:

\begin{enumerate}
    \item \textbf{Fast orientation must precede slow reasoning.} While classical tutoring favors extensive dialogue to elicit reasoning~\cite{Graesser2005}, students under severe time pressure actively resisted unskippable Socratic sequences. They did not reject deep learning as a goal; they wanted to decide \textit{when} deep learning was worth the time. Traditional ITS research has often treated shallow answer-seeking or repeated help requests as help misuse or ``gaming the system''~\cite{Aleven2003,Baker2004}. Our data complicate that interpretation: in a high-stakes homework environment, answer-first behavior can function as a rational self-regulation strategy for calibrating cognitive investment. One student summarized this distinction directly:
    \begin{quote}
    ``When time is tight, I do not have the mental space to study it deeply. I want the answer, plus a text version of the approach and a detailed explanation.'' (Chloe, formative interview)
    \end{quote}
    This became the rationale for a layered interface rather than a purely conversational tutor.

    \item \textbf{AI reduces the social friction of help-seeking.} Students described a split between school and home help-seeking. At school they often asked classmates or selected teachers; at home they turned to search and AI. Social embarrassment mattered. One student explained: ``I'm scared the teacher will mock me for asking a dumb question'' (Amy, formative interview). Another valued AI-style interruption because it avoided ``the awkwardness of raising your hand in class'' (Chloe, formative interview). This evidence reframed follow-up not as an optional chatbot feature, but as a private, low-stakes channel for asking the second or third question that students may suppress around adults or peers.

    \item \textbf{Curriculum trust and exam formatting are part of usability.} Students rejected AI solutions---even mathematically correct ones---if they used unfamiliar or out-of-syllabus methods. The educator participant strengthened this requirement from a classroom assessment perspective:
    \begin{quote}
    ``Our evaluation system is still Zhongkao and Gaokao, and ultimately it still has to land on paper-and-pencil problem solving.'' (educator participant)
    \end{quote}
    The same teacher emphasized that subject-specific answer norms can affect scores, including formula notation, table order, subscripts, and written explanations. This became the strongest justification for an \textit{Exam-Format Layer}: AITutor should keep explanations within junior-high syllabus scope and exam-conventional methods students can actually use in class and on the \textit{Zhongkao}.

    \item \textbf{Single-example understanding often becomes fake understanding.} Students and the educator both described a gap between understanding the current worked example and being able to solve a varied problem later. The teacher's practical test for real understanding was whether a student could handle changed conditions; if not, the original principle had not been internalized. This pushed the design beyond immediate answer pages toward delayed retrieval, wrong-book organization, and transfer practice.

    \item \textbf{Visual reasoning and local repair are prerequisites for hard problems.} Geometry and graph problems repeatedly failed when text introduced an auxiliary line, point, or relation that students could not visualize. Students asked for ``highlighter'' interactions that explain why the system made a particular move and for diagrams that reveal auxiliary lines step by step. This evidence connected the generative findings to two later features: step-linked visual grounding and contextual follow-up for local reasoning repair.
\end{enumerate}

\section{System Design and Usability Evolution}

Based on our generative findings, we designed the \textbf{AITutor} prototype. Rather than treating the AI as an open-ended chatbot, we constrained the interaction, translating pedagogical mechanisms into explicit UI features.

\newpage
\begin{figure}[H]
  \centering
  \begin{subfigure}[t]{0.5\columnwidth}
    \centering
    \includegraphics[height=2.08\linewidth,keepaspectratio]{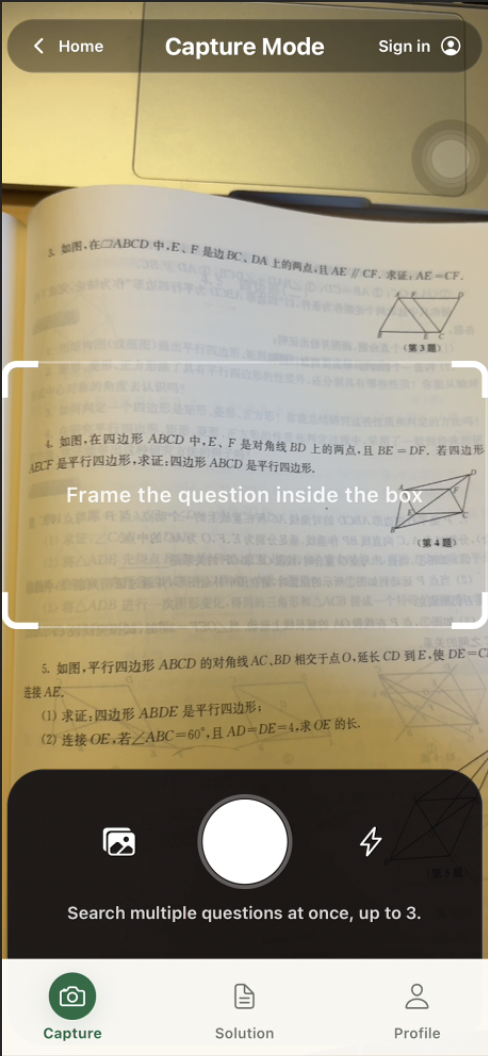}
    \caption{}
    \label{fig:capture}
  \end{subfigure}\hfill
  \begin{subfigure}[t]{0.5\columnwidth}
    \centering
    \includegraphics[height=2.08\linewidth,keepaspectratio]{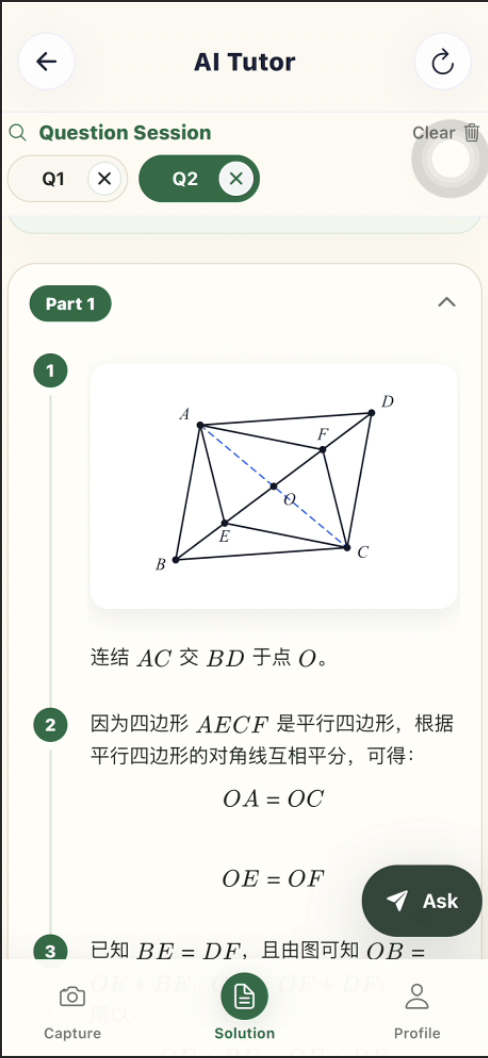}
    \caption{}
    \label{fig:solution_steps}
  \end{subfigure}
    \vspace{-4mm}
  \caption{AITutor's capture-to-solution flow. (a) Students photograph a homework problem. (b) The solution page presents numbered steps and links each step to the corresponding diagram element.}

  \label{fig:ui_capture_solution}
  \Description{Two AITutor screenshots: (a) the problem-capture camera and (b) a step-by-step solution with a synchronized geometric diagram.}
\end{figure}

\vspace{-2em}

\begin{figure}[H]
  \centering
  \begin{subfigure}[t]{0.48\columnwidth}
    \centering
    \includegraphics[height=2.08\linewidth,keepaspectratio]{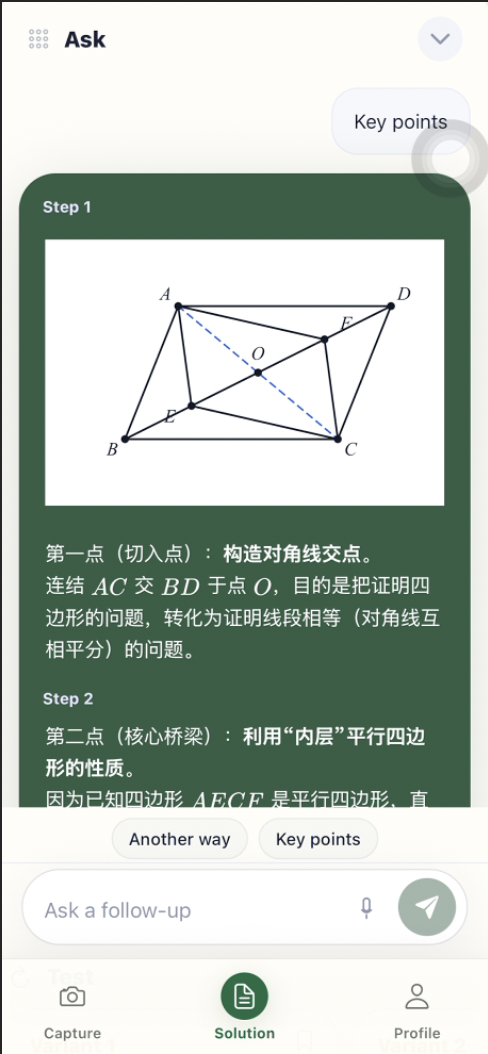}
    \caption{}
    \label{fig:followup_chat}
  \end{subfigure}\hfill
  \begin{subfigure}[t]{0.48\columnwidth}
    \centering
    \includegraphics[height=2.08\linewidth,keepaspectratio]{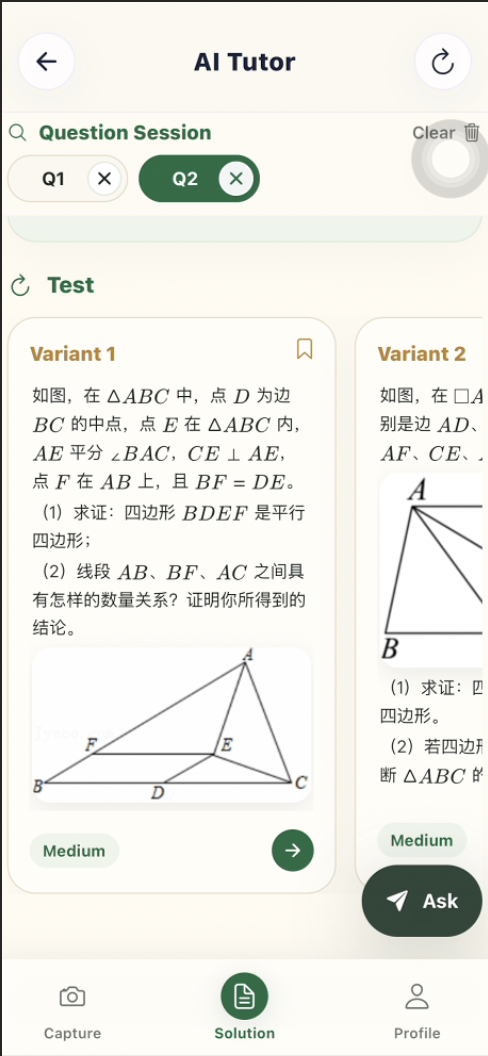}
    \caption{}
    \label{fig:transfer_practice}
  \end{subfigure}
  \vspace{-4mm}
  \caption{Interactive reasoning support. (a) Contextual follow-ups let students ask about a confusing step. (b) Transfer practice provides curriculum-matched variants for applying the same reasoning.}
  \label{fig:ui_interaction}
  \Description{Two AITutor screenshots: (a) contextual one-tap follow-up controls and (b) curriculum-matched transfer variants.}
\end{figure}

\begin{figure}[H]
  \centering
  \begin{subfigure}[t]{0.48\columnwidth}
    \centering
    \includegraphics[width=\linewidth]{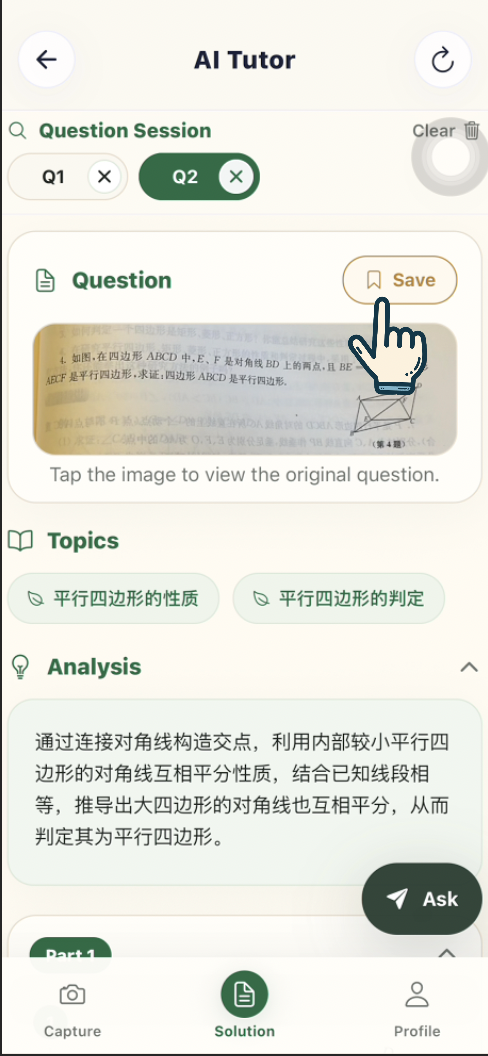}
    \caption{}
    \label{fig:solution_overview}
  \end{subfigure}\hfill
  \begin{subfigure}[t]{0.48\columnwidth}
    \centering
    \includegraphics[height=2.175\linewidth,trim=0 0 0 75bp,clip]{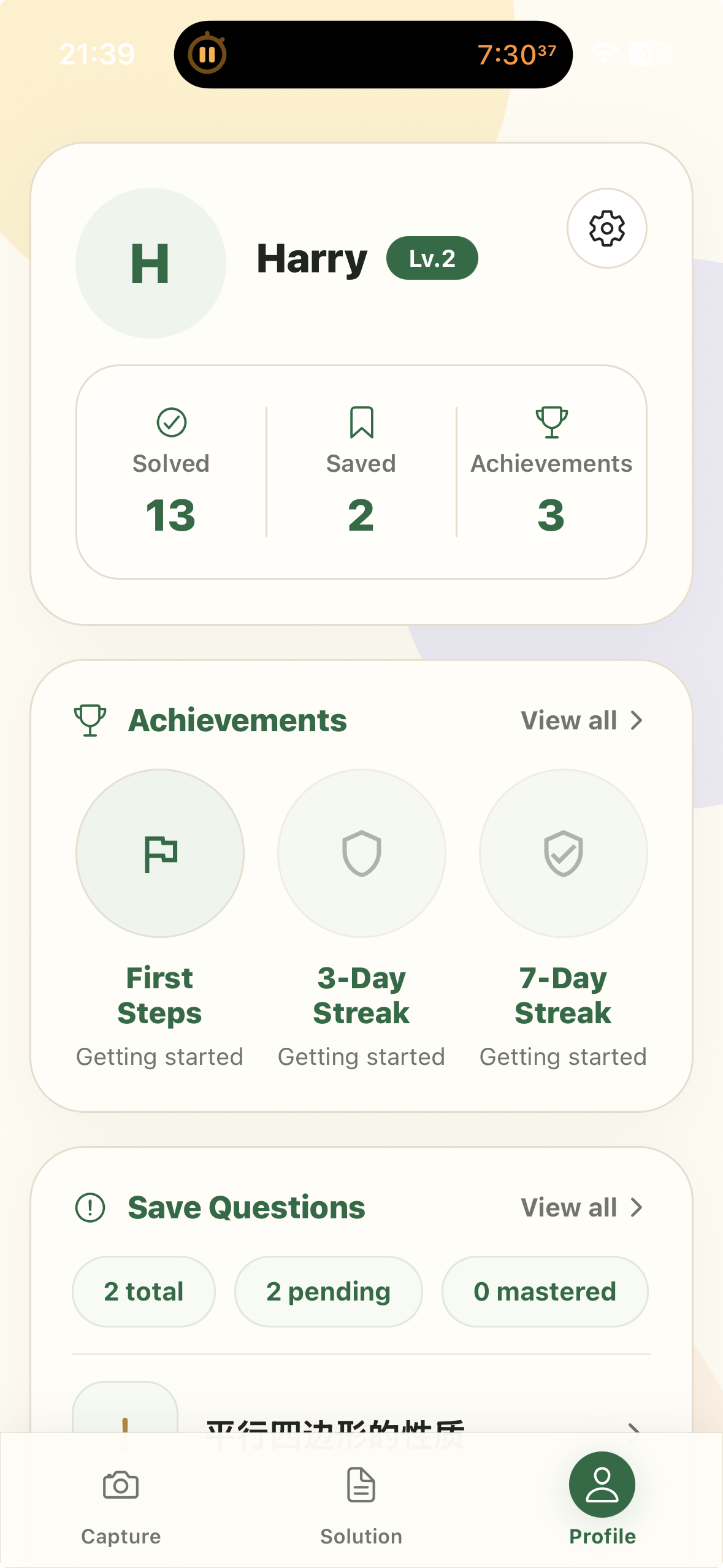}
    \caption{}
    \label{fig:saved_review}
  \end{subfigure}
  \vspace{-4mm}
  \caption{Curriculum fit and delayed review. (a) Students can save a recognized question for later. (b) The profile turns saved items into an automatic \textit{wrong-book} for delayed, spaced retrieval.}
  \label{fig:ui_review}
  \Description{Two AITutor screenshots: (a) the orientation view with the recognized question, knowledge points, and save/wrong-book controls and (b) the profile and wrong-book review page.}
\end{figure}

\subsection{Usability Study Insights and Design Rationale}
We conducted a usability study with 5 participants testing our initial prototype. Participants were junior-high or near-target students who regularly used paper homework, photo-search tools, or AI chat for math/science help. Each session used a think-aloud protocol around four tasks: capturing a problem, reading the generated solution, asking a follow-up about a confusing step or method, and saving or returning to a problem for later review. We combined observation notes with short post-task interviews, then grouped breakdowns by the underlying learning issue they exposed. The most important result was not a single usability bug, but a repeated pattern: students wanted help that was fast enough for homework while still preserving the reasoning mechanisms that make tutoring educationally valuable. Table~\ref{tab:usability_rationale} summarizes how observed breakdowns became design iterations.

\begin{table*}[t]
\centering
\caption{Usability breakdowns mapped to design rationale and system changes.}
\label{tab:usability_rationale}
\small
\begin{tabularx}{\textwidth}{>{\RaggedRight\arraybackslash}X >{\RaggedRight\arraybackslash}X >{\RaggedRight\arraybackslash}X}
\toprule
\textbf{Usability study breakdown} & \textbf{Underlying educational issue} & \textbf{System implementation} \\
\midrule
Participants faced a long ``thinking'' state and wanted to know the answer or core strategy before reading a full solution. & Homework pressure made students calibrate cognitive investment before committing to slow reasoning. & \textbf{Layered Worked Examples}: an overview (knowledge tags and core analysis) appears before labeled, expandable sub-question parts and step-by-step explanation; a final-answer checkpoint remains quickly accessible without forcing a long dialogue. \\
Step-by-step output was often procedurally dense but conceptually thin: students saw many steps without seeing why the key move was chosen. & More steps do not automatically create self-explanation; students need the hidden ``because, therefore'' link. & \textbf{Reasoning Layer}: steps are structured around what is done, why it is justified, and what it unlocks next. \\
Geometry explanations broke down when diagrams stayed blank, loaded late, or failed to show auxiliary lines as the text introduced them. & Text and figure were split across attention, increasing extraneous load for spatial reasoning. & \textbf{Step-Linked Visual Grounding}: the figure highlights points, labels, and auxiliary lines at the step where they matter. \\
Save/history features were valued, but students had to search for them and did not see how solving connected to later review. & Immediate completion and delayed learning were disconnected in the interface. & \textbf{Wrong-Book and Review Entry Points}: saving, history, and similar problems are reframed as delayed retrieval rather than optional storage. \\
Trust collapsed when the AI introduced cotangent (\textit{cot}) in a cube-projection task or vector methods in a coordinate-geometry task. & Mathematical correctness was insufficient if the method was outside junior-high classroom conventions; out-of-syllabus methods can also fail to earn process credit. & \textbf{Exam-Format Layer}: prompts and knowledge tags constrain outputs to junior-high syllabus scope and exam-conventional methods. \\
\bottomrule
\end{tabularx}
\end{table*}

The cotangent and vector incidents were especially consequential because they exposed a non-obvious form of trust failure. In both cases, students did not merely doubt whether the final answer was correct. They doubted whether the reasoning was \textit{usable} in their school context. This finding connected the usability study back to the related-work concept of calibrated trust: students needed evidence of curriculum fit before investing attention in the generated solution. The teacher interview explains why this is not just a preference issue. In China's paper-and-pencil exam ecology, a mathematically valid method can still be unusable if it crosses syllabus boundaries or skips answer-format conventions that determine process credit. It also sharpened our design stance on help-seeking cost. A generic chat box technically allowed students to repair the output, but requiring a student to type ``I have not learned vectors; please use junior-high knowledge'' is already too much cost under homework pressure. This is why the prototype shifted from a single long generated answer toward layered output, diagram-grounded steps, and step-level repair prompts.

\subsection{Resulting Pedagogical Interventions (AITutor UI)}

\subsubsection{Layered Worked Examples}
To mitigate cognitive overload and accommodate varied help-seeking paths, AITutor structures each solution in progressive layers (Figures~\ref{fig:capture} and~\ref{fig:solution_overview}). After capture, a \textit{Solution Overview} surfaces knowledge tags and a concise core analysis for curriculum-fit checking. An \textit{Orientation Layer} indexes sub-questions within a compound problem as labeled, expandable parts (e.g., Part~1, Part~2), so students can orient to the relevant sub-question before reading details. Expanding a part reveals the \textit{Explanation Layer}: step-by-step reasoning for that sub-question. A \textit{Final Answer} checkpoint remains accessible so students can compare their own result, then return to the relevant explanation layer. Across these layers, an \textit{Exam-Format Layer} keeps generated content within junior-high syllabus scope and exam-acceptable formats through knowledge tags, prompting, and curriculum-fit cues.

\subsubsection{Step-Linked Visual Grounding}
To support spatial reasoning, AITutor synchronizes textual steps with geometric diagrams (Figure~\ref{fig:solution_steps}). As the student scrolls through the text, corresponding auxiliary lines, angles, and points are highlighted. This concrete implementation of multimedia learning principles~\cite{Mayer2003} reduces the split-attention effect.

\subsubsection{Metacognitive Scaffolding}
Instead of relying on open-ended chat for follow-ups, AITutor provides a contextual follow-up dock: a floating action button opens a bottom sheet where students can tap suggested prompts (e.g., ``Key points,'' ``Another way'') or type a follow-up question. This lowers the interaction cost of help-seeking~\cite{Aleven2003} and scaffolds the student's metacognitive awareness (Figure~\ref{fig:followup_chat}).

\subsubsection{Delayed Retrieval and Review Loop}
Rather than adding practice burden during the current homework session, AITutor packages each solved problem into a delayed review loop: it offers curriculum-matched variant problems for optional transfer practice (Figure~\ref{fig:transfer_practice}), and organizes saved and solved items into a wrong-book review view for spaced weekend retrieval (Figure~\ref{fig:saved_review}).

\subsubsection{Implementation Note}
The deployed prototype was built as a React Native/Expo iPad app connected to a Python/FastAPI backend. The backend orchestrated LLM calls, curriculum-constrained prompting, retrieval over digitized course materials, and telemetry logging. The solution page used WebView-based math rendering for formulas and diagrams so that generated steps could remain readable in a mobile homework setting.

\section{Methodology}

To evaluate how students interacted with AITutor in authentic learning contexts, we deployed the system in the wild for a 12-day shared field window (May 15 to May 26).

\subsection{Participants and Study Window}
We recruited 12 junior-high students (P01--P12) from our target demographic (Grades 7--9). All 12 contributed quantitative telemetry logs. Ten students (P01--P10) participated in follow-up interviews, and eight of those ten also completed contextual observation sessions. P11 and P12 were telemetry-only participants because they were unavailable for qualitative sessions. Actual usage frequency varied naturally: some students used AITutor daily, while others concentrated their activity around weekends and homework-heavy periods. We use stable working IDs throughout the field-study analysis; where earlier formative participants had established pseudonyms, we retain them (Amy/P01 and Mia/P02; Table~\ref{tab:participants}).

\begin{table*}[t]
\centering
\caption{Participant profiles and data coverage for the field-study sample.}
\label{tab:participants}
\small
\begin{tabularx}{\textwidth}{>{\hsize=.05\hsize}X >{\hsize=.1\hsize}X >{\hsize=.05\hsize}X >{\hsize=.6\hsize}X >{\Centering\hsize=.1\hsize}X >{\Centering\hsize=.1\hsize}X}
\toprule
\textbf{ID} & \textbf{Alias} & \textbf{Gr.} & \textbf{Usage profile} & \textbf{Interview} & \textbf{Observation} \\
\midrule
P01 & Amy & 9 & Post-class/home review; re-explain missed classroom steps for wrong-problem digestion & Y & Y \\
P02 & Mia & 8 & Evening/weekend homework; visual step-by-step explanation for geometry/coordinate problems & Y & N \\
P03 & Lilly & 8 & Time-pressured evening homework; fast answer triage and rescue when stuck & Y & Y \\
P04 & -- & 8 & Homework-time difficult questions; stuck-problem and answer-confirmation support & Y & Y \\
P05 & -- & 8 & Wrong-problem correction and answer checking; detailed step explanation for learning & Y & Y \\
P06 & -- & 8 & Unsolved/important questions under time limits; quick answer-and-idea checks & Y & N \\
P07 & Mengmeng & 8 & After-homework difficult-problem batch review; detailed reasoning for hard geometry & Y & Y \\
P08 & PP & 7 & Homework/wrong-problem correction; follow-up and dynamic-visual explanation needs & Y & Y \\
P09 & KX & 8 & Post-homework hard problems and weekend review; history/saved-problem workflow & Y & Y \\
P10 & TYK & 9 & Answer confirmation on non-routine problems; trust-sensitive weakness diagnosis & Y & Y \\
P11 & -- & 8 & Telemetry-only participant; contributed instrumented app usage but no qualitative session & N & N \\
P12 & -- & 8 & Telemetry-only participant; contributed instrumented app usage but no qualitative session & N & N \\
\bottomrule
\end{tabularx}
\end{table*}

\subsection{Mixed-Methods Data Collection}
We employed a mixed-methods triangulation approach, combining instrumentation to capture \textit{what} students did at scale, with interviews and observation to surface \textit{why}.
\begin{itemize}
    \item \textbf{Instrumentation Logs:} We tracked 7,379 backend events across 104 valid sessions. We normalized timestamps to student local time, removed backend retries and known non-student traffic, constructed sessions using a 30-minute inactivity threshold, and filtered sessions with fewer than 3 events from funnel analysis. The logs monitored solve latencies, feature funnels, follow-up entry, transfer-card impressions, saved-practice creation, history access, and account-state transitions.
    \item \textbf{Contextual Observations (n=8):} We conducted 25--40 minute remote screen-share sessions during natural study routines in the evening or on weekends. Observers watched without intervening, taking structured field notes on problem selection, image capture/cropping, waiting behavior, result-page reading depth, diagram interaction, follow-up composition, save/history decisions, and exit moments.
    \item \textbf{Semi-Structured Interviews (n=10):} We conducted 20--30 minute retrospective interviews in Chinese. The protocol covered when students chose AITutor versus human help, how they selected submitted problems, how they read final answers versus steps, why they did or did not use follow-up/history/wrong-book features, and what specific incidents built or broke their trust. Participant quotations in this paper were translated from Chinese and lightly edited for readability.
\end{itemize}

\subsection{Data Analysis}
We analyzed qualitative data using a hybrid deductive-inductive approach. Three researchers independently coded the 10 interview transcripts and 8 observation field notes with descriptive labels such as answer-first checking, diagram-step mismatch, follow-up not noticed, and curriculum mismatch. We then conducted affinity diagramming to cluster incidents into reasoning-centered themes. A theme was retained only if it was supported by at least two participants and could be triangulated with another source, such as observation evidence or a quantitative funnel pattern.

\section{Field Study Findings}

Our field study reveals how students repurpose AI interfaces to manage cognitive load. We present our findings explicitly divided into quantitative behavioral patterns and qualitative thematic insights.

\subsection{Quantitative Findings}

\subsubsection{Overall System Usage and Latency}
\begin{figure}[H]
  \centering
  \includegraphics[width=\linewidth]{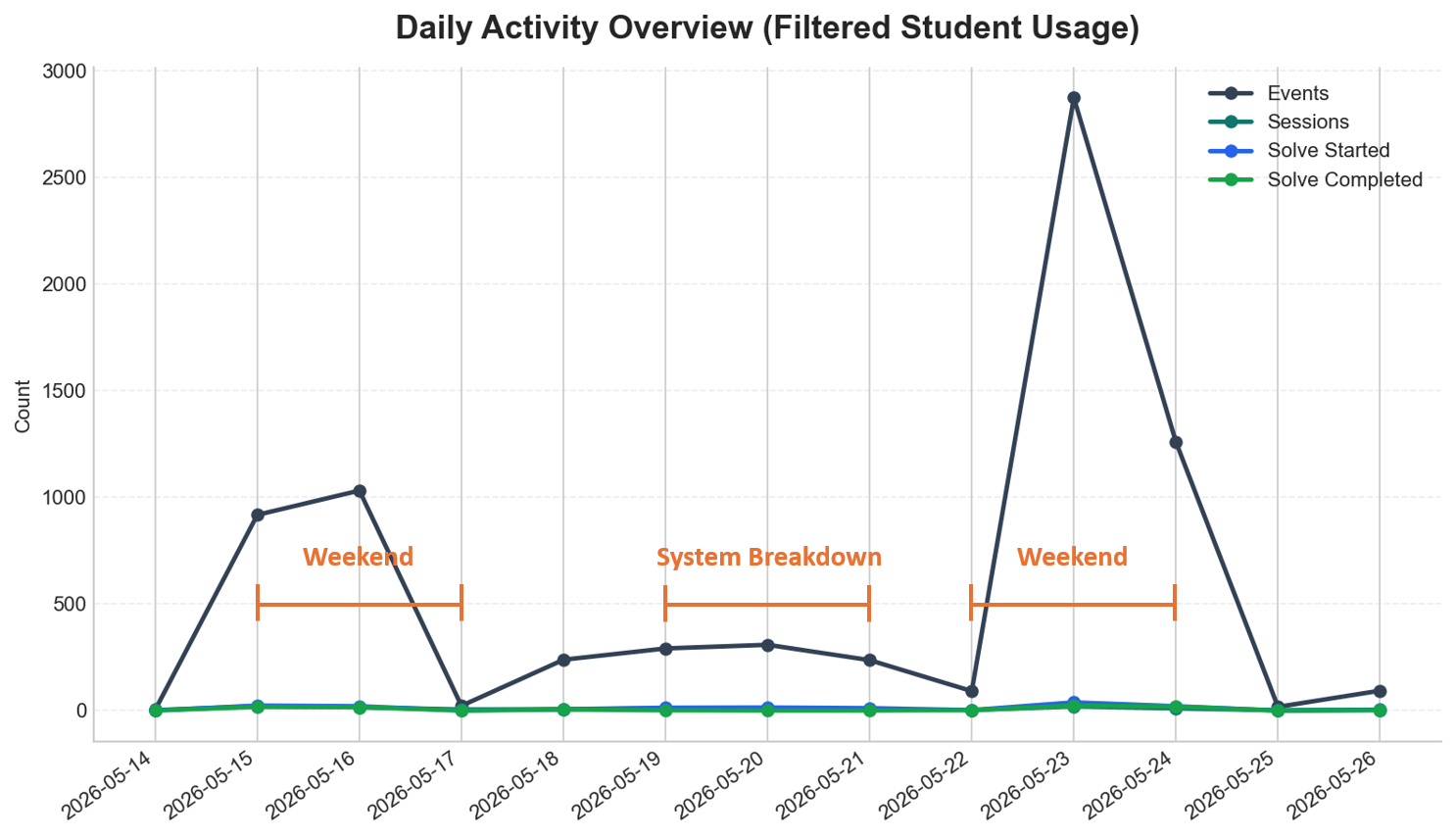}
  \caption{Daily solve outcomes. Activity spiked on weekends, indicating need-based rather than habitual usage.}
  \Description{Line chart of solve starts, completions, and failures across the field-study dates, with higher activity on Friday and weekend dates.}
  \label{fig:daily_outcomes}
\end{figure}

Overall system usage was meaningful but clustered around weekends and homework-heavy periods, demonstrating need-based engagement rather than a daily habit (Figure~\ref{fig:daily_outcomes}). This temporal pattern reinforced the central tension of the project: students did not approach AITutor as a calm daily learning environment, but as a just-in-time reasoning aid during homework bursts.

\begin{figure}[H]
  \centering
  \includegraphics[width=\linewidth]{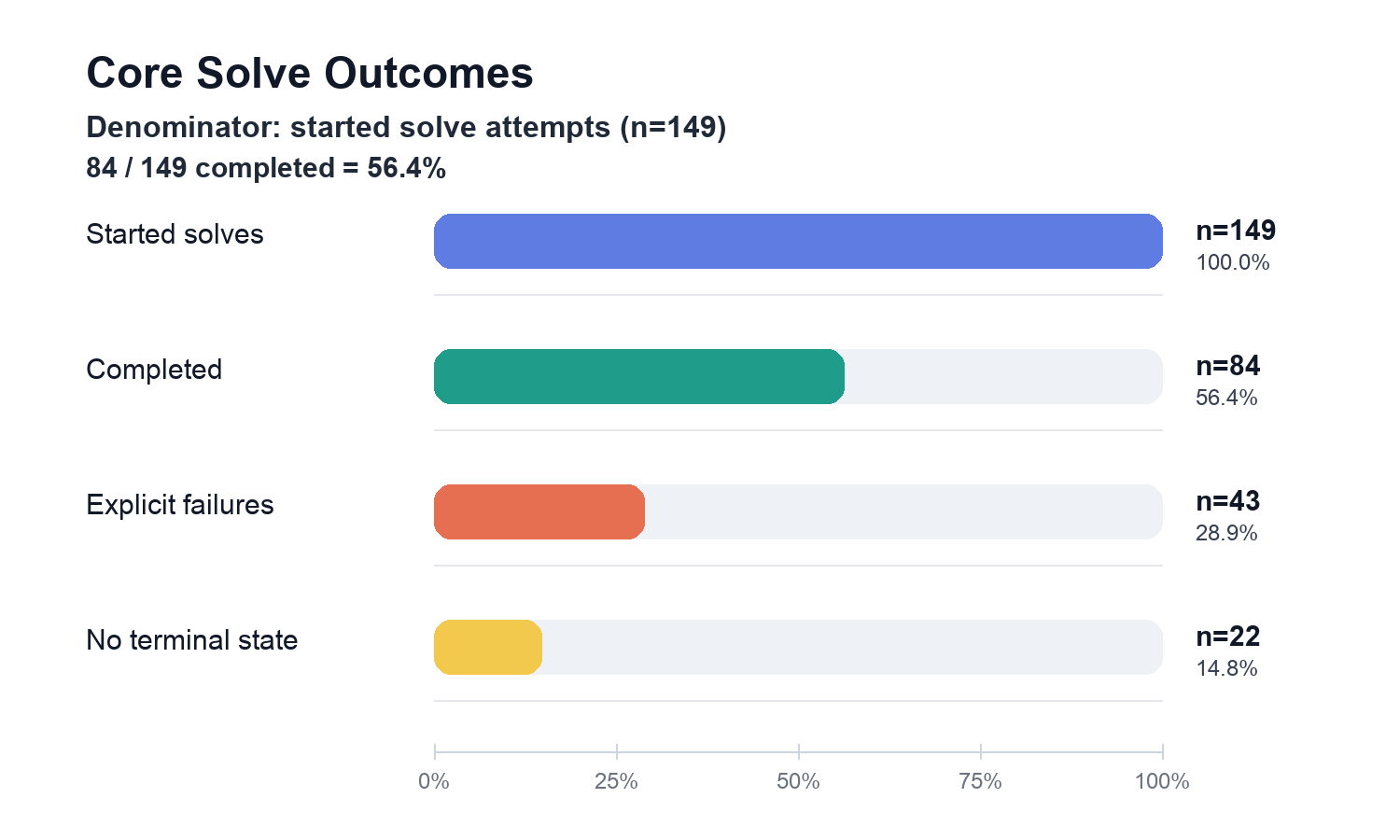}
  \caption{Core solve outcomes using started solve attempts as denominator (n=149). 84 solves completed successfully (56.4\%); 43 failed explicitly and 22 did not reach a completed terminal state.}
  \Description{Funnel chart using 149 started solve attempts as the denominator, showing 84 completed successfully, 43 explicit failures, and 22 incomplete or non-terminal attempts.}
  \label{fig:solve_funnel}
\end{figure}

The core solve flow revealed a significant bottleneck. To avoid denominator confusion, Figure~\ref{fig:solve_funnel} uses started solve attempts as the denominator: students started 149 solve attempts, and 84 completed successfully, yielding a 56.4\% completion rate. This number should not be interpreted as simple user disinterest. During the shared field window, the system experienced a reliability breakdown on May 20 and May 21, during which no started solves completed; May 19 was also unusually weak, with only 1 of 13 started solves completing. Thus, the funnel combines user behavior with a system-stability event. Latency telemetry further explains abandonment: the average solve latency was 32.0 seconds, with the p90 reaching 68.5 seconds. System stability and prolonged wait times became a prerequisite bottleneck, directly disrupting students' cognitive momentum and causing premature abandonment.

\subsubsection{Feature Funnels: Follow-Up and Transfer Practice}
\begin{figure*}[t]
  \centering
  \begin{minipage}[t]{0.47\textwidth}
      \centering
      \includegraphics[width=\linewidth]{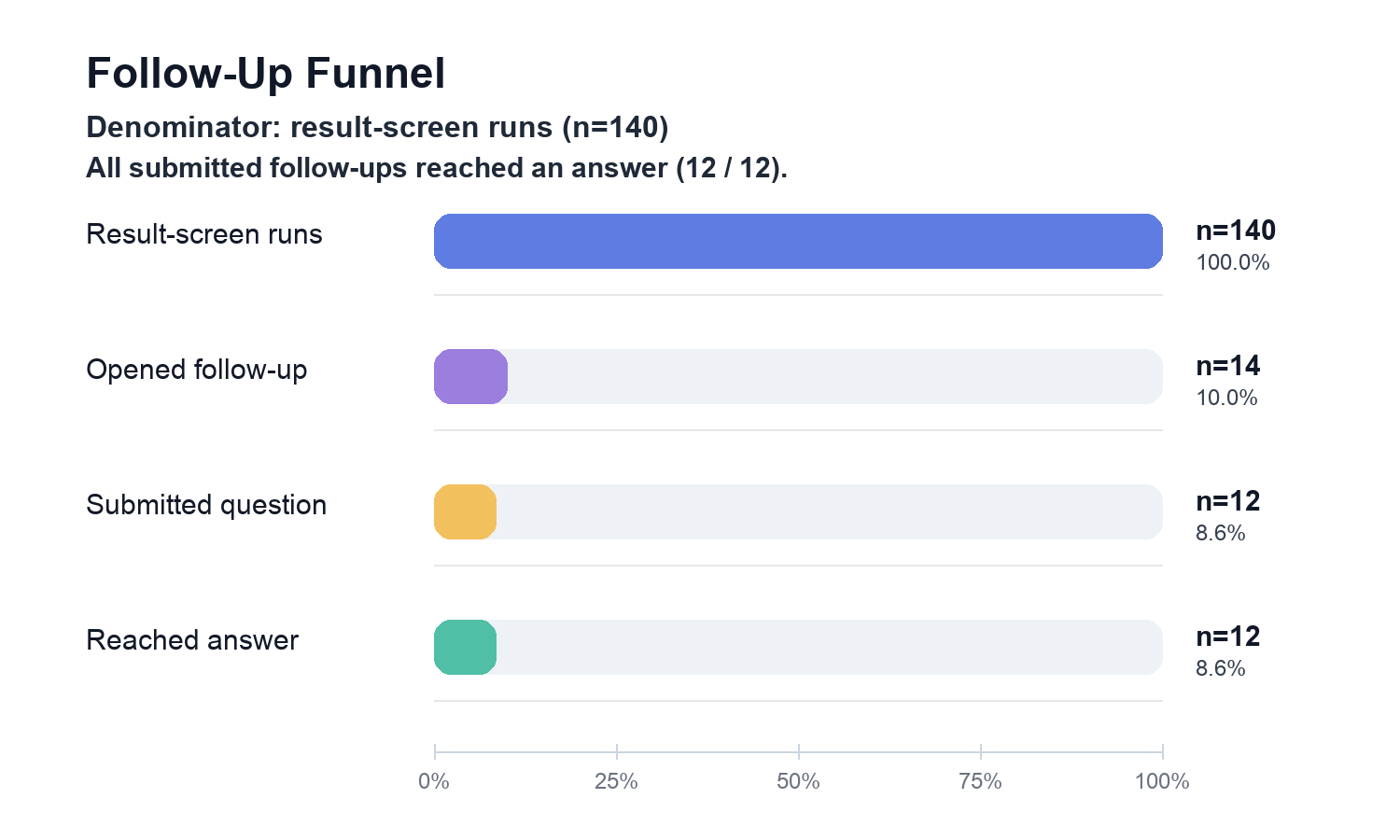}
      \caption{Follow-up funnel using result-screen runs as denominator (n=140). Students opened follow-up 14 times (10.0\%), submitted 12 questions (8.6\%), and all 12 reached an answer.}
      \label{fig:followup_funnel}
  \end{minipage}\hfill
  \begin{minipage}[t]{0.47\textwidth}
      \centering
	      \includegraphics[width=\linewidth]{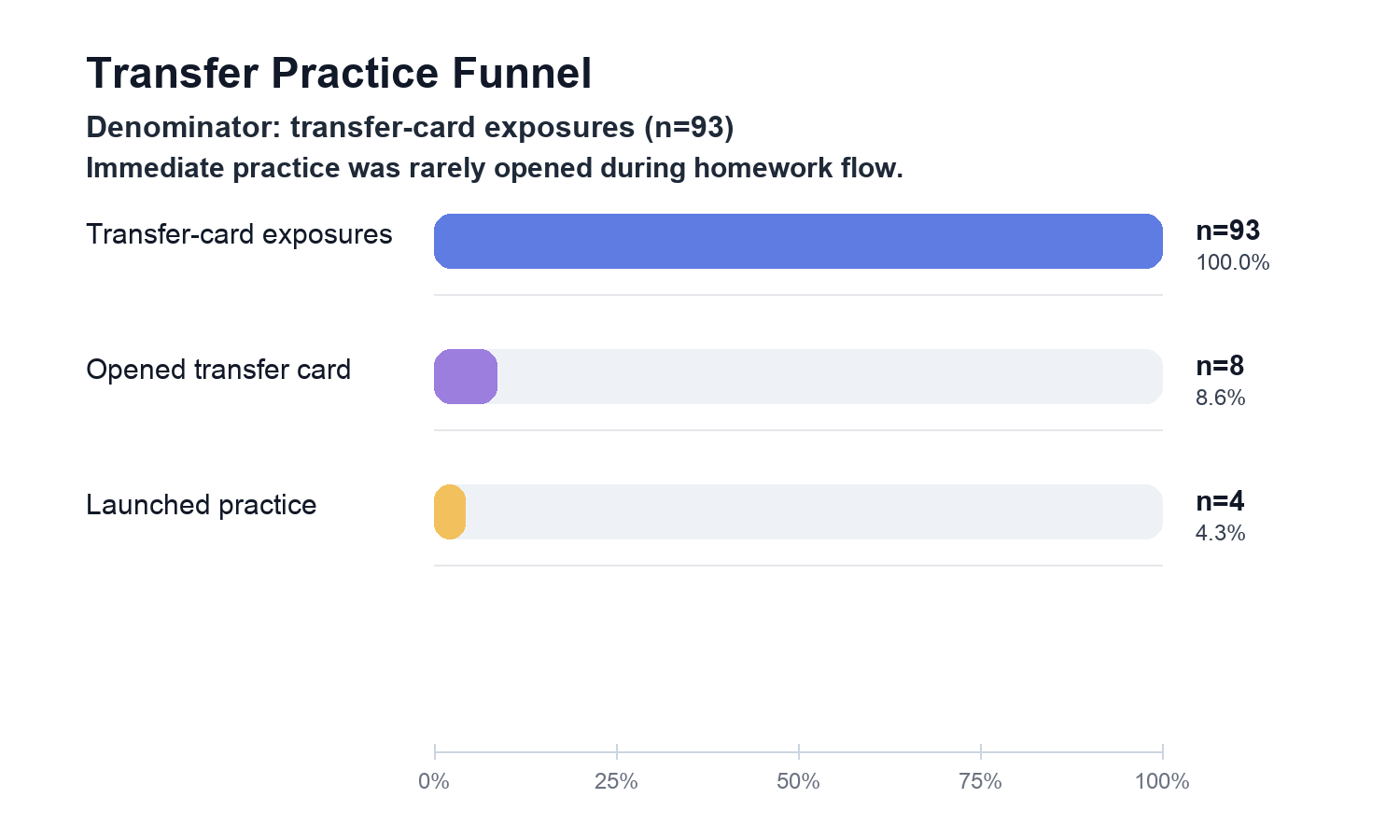}
	      \caption{Transfer funnel using transfer-card exposures as denominator (n=93). Students opened 8 cards (8.6\%) and launched 4 practice questions (4.3\%).}
	      \label{fig:transfer_funnel}
	  \end{minipage}
  \Description{Two funnel charts showing low follow-up initiation and low transfer-practice opening after result screens.}
\end{figure*}

Telemetry showed a low overall entry rate for follow-up questions. Across 140 result-screen runs, students opened the follow-up interface 14 times (10.0\%) and submitted 12 follow-up questions (8.6\%, Figure~\ref{fig:followup_funnel}). However, once a follow-up question was submitted, all 12 reached first content. This means the feature functioned reliably; the bottleneck was initiation.

Similarly, the similar-question transfer feature appeared frequently after completed solves, but students rarely chose to open it immediately (Figure~\ref{fig:transfer_funnel}). We observed 93 transfer-card exposures; students opened 8 cards (8.6\% of exposures) and launched 4 practice questions (4.3\% of exposures). This is a strong example of a timing mismatch: transfer practice may be educationally valuable, but it conflicts with the immediate goal of finishing the current homework task.

Account-state telemetry exposed a second form of friction. Across all filtered events, 5,155 events (69.9\%) occurred in guest mode, while only 2,224 events (30.1\%) came from logged-in users. This suggests that students were willing to ask for lightweight, one-off help without committing to an account, even though login enabled the history and saved-practice features needed for long-term review. The data reveal a product tension between \textit{lightweight help} and \textit{durable learning records}; this motivated our later emphasis on automated wrong-book generation and clearer value for account-backed review.

\subsection{Qualitative Findings}

Table~\ref{tab:joy-frustration} summarizes representative joy and frustration incidents that ground the qualitative themes below.

\begin{table*}[!t]
\caption{Joy and frustration incidents observed during the field study.}
\label{tab:joy-frustration}
\small
\begin{tabularx}{\textwidth}{>{\hsize=.2\hsize\RaggedRight\arraybackslash}X >{\hsize=.45\hsize\RaggedRight\arraybackslash}X >{\hsize=.35\hsize\RaggedRight\arraybackslash}X}
\toprule
\textbf{Experience type} & \textbf{Specific incident} & \textbf{Why it matters} \\
\midrule
Joy: understanding the missed step & Amy returned to a problem that the teacher had explained but that still contained a confusing middle step. AITutor helped when it broke open that missing step (P01). & AITutor's value is strongest when it acts like a patient post-class tutor, not just an answer generator. \\
Joy: finding the exact mistake & Lilly had already set up a function-graph problem correctly but made a calculation error. Seeing the final answer first helped her identify where her work went wrong (P03). & For answer-first students, showing the final answer early can support self-diagnosis rather than simple copying. \\
Joy: making geometry visible & Students valued geometry explanations where diagrams, auxiliary lines, and dynamic cases were visible rather than described only in text (P02, P07, P08, P09). & Visual reasoning is a real differentiator from ordinary homework-search products. \\
Frustration: solving the wrong question & One student framed question 5 on a worksheet, but the app solved question 4; another worried that extra captured content would cause the same error (P05, P06). & Recognition errors damage trust before explanation quality can matter. \\
Frustration: waiting under pressure & Some students felt the app was slow after 20-30 seconds without useful text, and switched to another activity (P02, P03, P06). & Speed is partly technical, but it is also interaction design: students need safe parallel action. \\
\bottomrule
\end{tabularx}
\end{table*}

\subsubsection{Theme 1: Layered Explanations Turn Answers into Reasoning Checkpoints}
A counter-intuitive finding emerged regarding "answer-first" behavior. While often stigmatized in education as shortcutting, checking the final answer first served as a critical orientation device. Traditional ITS research has often interpreted answer-seeking and shallow help use as ``gaming the system''~\cite{Baker2004}. However, our data suggest that in high-stakes environments, answer-first behavior can function as a rational self-regulation strategy for calibrating cognitive investment. Lilly (P03) used the final answer to diagnose whether her error in a function-graph problem was a conceptual setup mistake or a mere arithmetic calculation error.
\begin{quote}
``I look at the final answer first. If it matches, then I read the steps below.'' (P03, interview)
\end{quote}
By confirming the answer first, students calibrated their cognitive investment before reading the detailed steps. This pattern aligns with self-explanation and feedback research: a worked solution can support learning when it helps the student compare their own reasoning against a meaningful target and decide what to inspect next~\cite{Chi1989,Hattie2007,Bisra2018}. Conversely, other students (e.g., P01) completely bypassed the final answer to hunt for one specific missing intermediate classroom step. This validates our layered design: different reasoning gaps require different entry points into the worked example. The design contribution is therefore not "show the answer earlier" in isolation, but use the answer as a checkpoint that routes students into the right depth of reasoning.

\subsubsection{Theme 2: Step-Linked Visualization Externalizes Reasoning}
For geometry and dynamic-point problems, diagrams functioned as the core reasoning engine. Mia (P02) noted that the tool's ability to draw auxiliary lines synchronously with the text significantly reduced the effort of matching proof lines to the drawing:
\begin{quote}
``Homework-help apps sometimes only show one diagram and a pile of steps underneath. I have to keep scrolling back to the figure. If the figure changes with the text and the lines, angles, and points appear step by step, I do not have to scroll back and forth.'' (P02, interview)
\end{quote}
When the system failed to visually anchor a label (e.g., introducing a point $P1$ in text without displaying it), students' reasoning immediately broke down. This mirrors multiple-representation research: visual support must be coordinated with the symbolic or textual step it is meant to explain~\cite{Ainsworth1999,VanDerMeij2006}. Visualization is not a decorative bonus; it is the scaffolding for spatial cognition.

\subsubsection{Theme 3: Follow-Up Works as Local Reasoning Repair}
While quantitative data showed low follow-up entry rates, interviews revealed that when used, follow-ups provided immense value for \textit{local reasoning repair}. Across 140 result-screen runs, only about 10\% entered the follow-up flow; however, once students submitted a follow-up, 100\% reached first content and completed successfully. Amy (P01) used the feature to clarify a single confusing substep in a proof. Students expressed a strong preference for localized, contextual follow-ups rather than general chat. This extends classic scaffolding and cognitive apprenticeship ideas: support should make the next expert move visible at the point where the learner cannot proceed alone~\cite{Wood1976,Collins1989}. The bottleneck was not whether AI could answer follow-up; it was the friction of noticing the feature and composing a precise metacognitive question under time pressure. P03 explained this failure mode directly:
\begin{quote}
``If a problem is too hard, I just give up. I am not used to typing out my question again.'' (P03, interview)
\end{quote}
This is why one-tap repair buttons are theoretically important: they reduce the cost of help-seeking without removing the need to inspect the local reasoning step~\cite{Aleven2003,Roll2011}.

\subsubsection{Theme 4: Delayed Reasoning Loops Over Immediate Practice}
\begin{figure}[H]
  \centering
  \includegraphics[width=\linewidth]{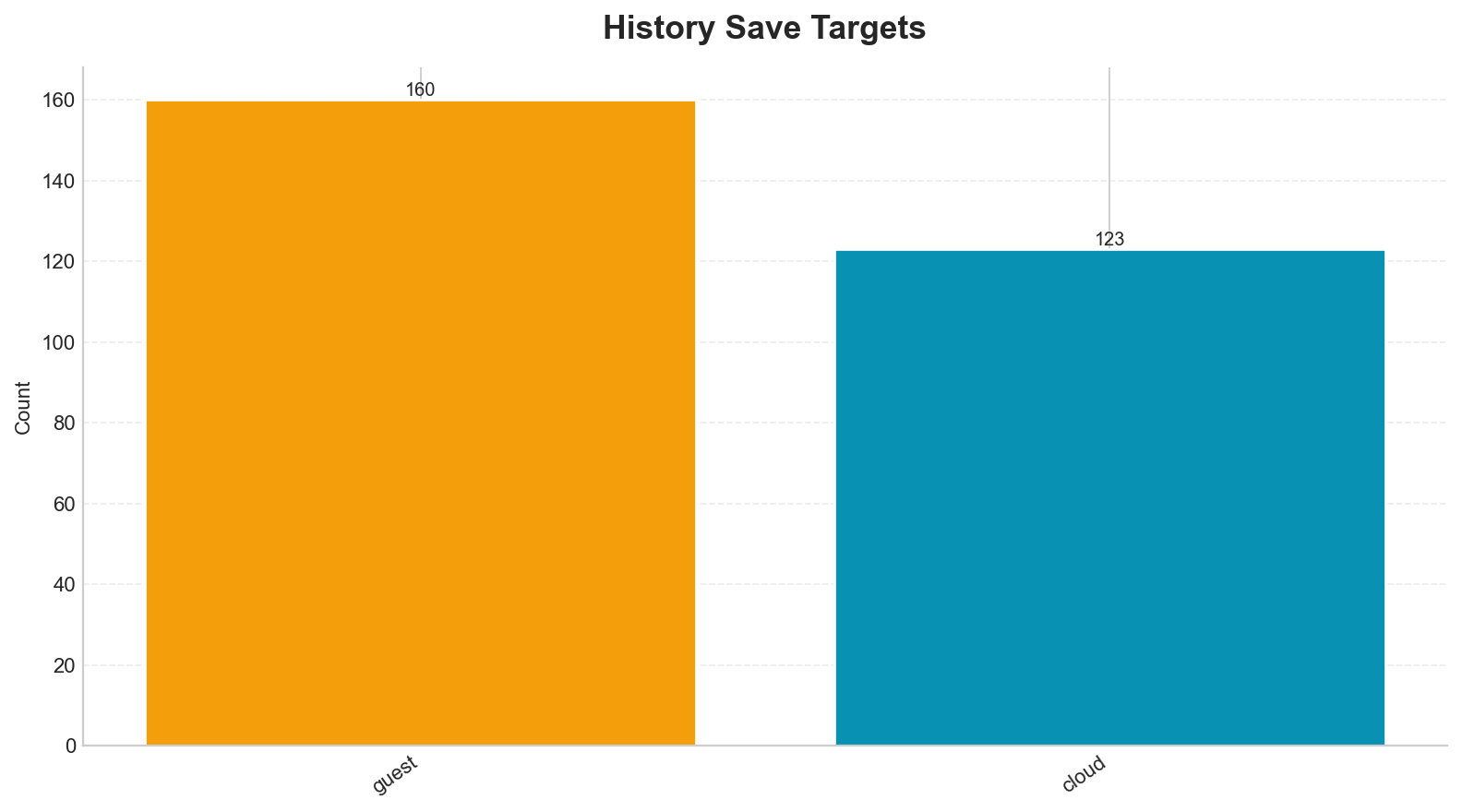}
  \caption{History save targets. Students proactively saved difficult items, highlighting the demand for delayed retrieval workflows.}
  \Description{Bar chart summarizing the types of problems students saved, emphasizing difficult or uncertain items for later review.}
  \label{fig:history_save_targets}
\end{figure}

As Figure~\ref{fig:transfer_funnel} illustrates, students rarely opened transfer cards immediately after completing a solve. Qualitative data revealed this was a timing issue, not a lack of perceived value.
\begin{quote}
``I can barely finish the homework. How would I have time to do similar problems? Just seeing them feels overwhelming.'' (P03, interview)
\end{quote}
Students like KX (P09) and TYK (P10) still wanted AITutor to organize these problems into a knowledge-map "wrong-book" for delayed retrieval practice during weekend reviews. This matches the spacing and retrieval literature: practice is more useful when it reactivates knowledge later rather than piling extra work onto the current task~\cite{Cepeda2006,Dunlosky2013}. Telemetry corroborates this intent: students selectively saved problems to their history based on difficulty (Figure~\ref{fig:history_save_targets}), validating the need for automated wrong-book organization to support spaced retrieval and structurally meaningful transfer~\cite{PanRickard2018}.

\subsubsection{Theme 5: Curriculum Fit Determines Trust Investment}
Students treated AI trust not as a measure of mathematical absolute truth, but as a measure of curriculum alignment. When KX (P09) noticed the AI using high-school knowledge for a junior-high problem, she immediately abandoned the solution:
\begin{quote}
``It used a high-school knowledge point. After I saw that I had not learned it, I stopped reading.'' (P09, interview)
\end{quote}
This echoed the usability study, where one participant lost confidence after the system introduced cotangent (\textit{cot}), and another rejected a vector-based method until the system was asked to use junior-high knowledge. Students demanded explicit trust signals: recognized problem regions, extracted conditions, and "junior-high method" tags. This supports a calibrated-trust interpretation: the interface should help students decide when to rely on AI rather than simply increasing confidence in every output~\cite{Lee2004,Kizilcec2016,Bucinca2021}. For this population, calibrated trust is not only about whether the AI computes correctly; it is about whether the method is learnable, writeable, and acceptable in the student's exam ecology. Without these signals, students refused to invest the cognitive effort required to read the reasoning chain.

\section{Discussion and Implications for Design}

Our study suggests that AITutor was useful in daily life as a just-in-time reasoning aid rather than as a daily learning companion. It fit into students' existing ecosystem between fast photo-search products, human teachers, tutoring classes, and handwritten wrong books: students used it when they needed a clearer explanation than search results but could not or did not want to ask a person. This interpretation also connects back to the literature. AITutor's value came from making LLM output more step-granular, visually grounded, locally repairable, and recoverable for later retrieval, which maps directly onto prior work on interaction granularity, multimedia learning, help-seeking cost, and delayed practice.

\subsection{Prioritized Improvements}
Based on our field study findings and participants' explicit requests, we identified five critical improvements required to transition the prototype into a robust pedagogical tool:
\begin{enumerate}
    \item \textbf{Faster Access to Core Strategy:} Introduce a dual-mode explanation structure where the "Answer + Core Idea" layer loads instantly, reducing the 30-second latency bottleneck that causes students to abandon the app.
    \item \textbf{Stricter Curriculum Alignment:} Implement explicit tags that verify whether a generated method belongs to the junior-high syllabus, preventing the use of advanced (e.g., vector-based or calculus) methods that students cannot use in exams.
    \item \textbf{Dynamic Geometry Coordination:} Require the UI to automatically highlight or draw auxiliary lines directly on the provided geometric diagram synchronously with the textual explanation steps.
    \item \textbf{Step-Specific Follow-Up Buttons:} Replace the generic follow-up dock at the bottom of the screen with localized, step-specific buttons (e.g., "Explain this step," "Simpler method") to minimize interaction friction.
    \item \textbf{Automated Wrong-Book Generation:} Automatically segment captured problems and organize them by knowledge point into a delayed-retrieval review list, transforming immediate "transfer tasks" into spaced weekend practice.
\end{enumerate}

\subsection{Implications for the Broader AI Education Domain}
Beyond specific app features, our study provides a broader theoretical framework for educational AI, which we term the \textbf{Reasoning-Centered Product Loop}:
\begin{itemize}
    \item \textbf{Orient: Calibrating Cognitive Investment.} Educational AI should not hide answers under the guise of preventing cheating. In time-pressured settings, the final answer can help students decide whether to invest effort in self-explanation, error search, or full solution reading~\cite{Chi1989,Hattie2007}. The design question is how to make answer access an entry point into reasoning rather than an endpoint.
    \item \textbf{Visualize: Coordinating Mental Models Across Representations.} Multimodal AI must treat text and diagrams as a single synchronized reasoning object. Unlinked generative text creates split-attention effects for spatial reasoning~\cite{Mayer2003,VanDerMeij2006}; step-linked diagrams can reduce this burden by making the relevant relation visible at the moment it is invoked.
    \item \textbf{Repair: Reducing the Metacognitive Friction of Help-Seeking.} Generic conversational chatbots are ill-suited for time-pressured homework because students must first know how to ask a good metacognitive question. The domain should shift toward contextual, step-specific repair buttons (e.g., "Explain this step," "Use junior-high method") that lower the cost of help-seeking while preserving the learner's attention on the local reasoning breakdown~\cite{Wood1976,Aleven2003,Roll2011}.
    \item \textbf{Retrieve: Turning Solutions into Delayed Reasoning Loops.} Learning interfaces must shift "similar problem" recommendations away from the immediate homework session, where they are viewed as a burden, and inject them into delayed, spaced-retrieval review workflows~\cite{Roediger2006,Cepeda2006,PanRickard2018}. The product goal is not simply to store solved questions, but to bring back the original reasoning relation when the student is ready to practice.
    \item \textbf{Verify: Reframing Calibrated Trust as Curriculum Fit.} Because students decide whether to read a solution based on exam usability, AI tutors should expose recognized problem regions, extracted conditions, knowledge points, and grade-appropriate method labels before asking students to trust a generated reasoning chain~\cite{Lee2004,Bucinca2021,Kasneci2023}. Our study extends calibrated-trust work by showing that, for exam-oriented learners, distrust often targets the \textit{method's legitimacy} rather than only the answer's correctness.
\end{itemize}

\section{Limitations and Future Work}
This study has several limitations that should shape interpretation and future work. First, the field deployment lasted only 12 calendar days. Students may have experienced a novelty or Hawthorne effect: they knew they were trying a new class prototype, and some usage may reflect curiosity rather than stable long-term learning behavior. A longer deployment across multiple homework cycles and exam-review periods is needed to evaluate retention, wrong-book reuse, and whether answer-first checking continues to function as a reasoning checkpoint over time.

Second, the sample size was small. All 12 participants contributed telemetry, but only 10 completed interviews and 8 completed contextual observations. This was sufficient for formative HCI analysis and triangulation, but it does not support population-level claims about Chinese junior-high students. Future studies should include a larger and more diverse sample across regions, school types, and achievement levels, and should distinguish students who use AI primarily for shortcutting from those who use it for self-regulated repair.

Third, the telemetry was affected by an unavoidable system-reliability incident. On May 20 and May 21, no started solves completed, and this outage directly depressed the 56.4\% solve-completion funnel. We therefore treat solve completion as a combined measure of user behavior and system stability, not as a pure engagement metric. Future deployments should separate technical availability from user abandonment more rigorously and instrument partial-output states so that students can continue reasoning while a full solution is still loading.

Finally, our current evidence focuses on interaction behavior and perceived reasoning support rather than measured learning gains. Future work should add pre/post assessments, delayed transfer tasks, and log-linked analysis of whether students can redo saved wrong problems after spaced review. This would test whether the Reasoning-Centered Product Loop improves durable mathematical understanding rather than only making help-seeking feel more usable.

\section{Conclusion: Educational Equity Through HCI}

By translating learning-science theories into concrete UI interventions, we demonstrated how AI tutoring systems can preserve and enhance mathematical reasoning. Moving forward, HCI design in education must prioritize the scaffolding of the reasoning process over the mere fluency of the generated solution. AITutor illustrates that the future of educational AI lies not in answering questions better, but in designing interfaces that teach students how to think about them.

Crucially, this reasoning-centric approach has profound implications for \textbf{educational equity}. Generative AI holds the promise of democratizing access to high-quality tutoring. However, if AI tools remain simple "answer engines," they risk widening the gap between students who naturally possess high metacognitive regulation (who use AI to verify and learn) and those who do not (who use AI to bypass cognitive struggle). Our guest-mode telemetry makes this equity issue concrete: 69.9\% of filtered events occurred without login, showing that students often remain in lightweight, one-off help mode even when long-term learning requires records, classification, and return visits. For well-resourced students, a human tutor often performs that metacognitive labor by curating wrong problems and deciding what to revisit. For students without that support, requiring manual registration, saving, tagging, and review planning can keep them stuck at shallow answer lookup. Automated wrong-book generation is therefore not only a convenience feature; it lowers the cognitive and operational barrier to long-term learning.

In highly competitive education systems like China's \textit{Zhongkao} environment, students with abundant resources rely on expensive, highly responsive human tutors to provide this exact metacognitive scaffolding. A human tutor observes a student's hesitation, dynamically sketches an auxiliary line, pauses to ask a guiding question, and curates a customized list of wrong problems for weekend review. By embedding step-linked visual grounding, contextual follow-up for reasoning repair, and automated spaced retrieval directly into the AITutor interface, we provide students who lack access to such resources with the same structured, patient, and rigorous pedagogical guidance. Ultimately, educational equity through AI is not achieved merely by lowering the cost of answers, but by democratizing access to robust reasoning processes.

\bibliographystyle{ACM-Reference-Format}
\bibliography{sample-base}
\end{document}